\documentclass[referee]{raa}

\usepackage{graphicx,times}
\usepackage{natbib}
\usepackage{amssymb,amsmath}
\bibpunct{(}{)}{;}{a}{}{,}
\usepackage{xcolor}

\usepackage[pagebackref=true]{hyperref}

\begin{document}

   \title{Outflow from unmagnetized shocked radiative transonic
   accretion disk around a black hole}

 \volnopage{ {\bf 20XX} Vol.\ {\bf X} No. {\bf XX}, 000--000}
   \setcounter{page}{1}

   \author{Arghya Chaudhuri \inst{1}, 
   Apurba Ghosh\inst{1,2}, Sudip K Garain\inst{1,2,*} }

   \institute{ Department of Physical Sciences, Indian Institute 
   of Science Education and Research Kolkata, Mohanpur - 741 246, 
   WB, India; {\it sgarain@iiserkol.ac.in}\\
        \and
   Center of Excellence in Space Sciences India, Indian Institute 
   of Science Education and Research Kolkata, Mohanpur - 741 246, 
   WB, India\\
\vs \no
   {\small Received 20XX Month Day; accepted 20XX Month Day}
}

\abstract{We study outflow from an unmagnetized, shocked accretion disk
around a non-rotating super-massive black hole using
multidimensional hydrodynamics simulation with radiative cooling.
We aim to investigate whether such shocked accretion flow can
launch sustained collimated bipolar outflow reaching out to thousands
of gravitational radii even in the absence of magnetic field
and if yes, what terminal velocity
can they achieve? We present the results of a few simulations of
geometrically thick accretion flow with increasing specific angular momentum
on a vertically elongated cylindrical domain. We show that
bipolar outflow from a region very close to the black hole is
originating and propagating vertically out to our simulation
domain boundary at around $2651$ Schwarzschild
radius. The outflow attains a terminal velocity with a maximum
value found to be $0.14c$ and the outflow rate depends on
the angular momentum value of the accreting material.
We also compute the self-Comptonized bremsstrahlung spectra for
all the disk-jet runs.
}

   \authorrunning{Chaudhuri et al. }            %author_head in even pages
   \titlerunning{Outflow from shocked accretion}  % title_head in odd pages
   \maketitle

%________________________________________________ sections below
%
\section{Introduction}           %% first-level sections will be auto-capitalized
\label{sec:intro}

Many high-energy sources containing black hole show wind and/or jet
\citep{mirabel1994,cygjet2001,gallo2010,blan2019}.
Whether the black hole is part of a binary system
or located at the core of a galaxy, the qualitative structure of outflow
is very similar when it is present. The outflow originates
from very close to the central object and extends from hundreds to
thousands of gravitational radii. It is well understood that,
unlike a normal star, a black hole itself does not eject matter.
Rather, the gravitationally captured matter, forming an
accretion disk around the black hole, produces the outflow.
Given that the gravitational attraction is very strong close
to a black hole, it is thus very surprising how a fraction
of the infalling matter gets rid of this strong attraction and
becomes unbounded.

Outflow formation from the magnetized plasma of an accretion
disk is mainly attributed to two mechanisms:
Blandford-Znayek \citep{bz1977} and Blandford-Pyne \citep{bp1982}.
It is, of course not very surprising that the accreting
material is magnetized. Magnetic field is very pervasive and wanders
along with plasma. In a binary system, matter can drag
in magnetic field from the companion star. In the case of accretion
onto a galaxy center, the infalling matter can drag magnetic field
from nearby stars or from the interstellar medium.
As this magnetized matter moves closer
to the black hole, the magnetic field structure, as well as the
magnitude may change due to various physical processes such as
geometric compression, dynamo effect, magnetic reconnection, etc.
This magnetic field can then help in shaping the accretion disk
structure and launching outflow from the
accretion disk in the manner described in these references.

However, the magnetic field may not be the only driving force behind
launching outflow. In principle, if a sufficiently strong force
develops in a way that cancels the gravitational pull, matter can be
driven away from the accretion disk. Thus, if the outward centrifugal
force due to rotating matter becomes sufficiently strong, the effective
potential barrier \citep{gravi,st1983} may be so high that the matter may
not be able to accrete onto the central black hole and
be driven away.
Also, a sufficiently strong pressure gradient
force acting opposite to the gravitational attraction can drive
material away. The pressure can be due to radiation or thermal.
If the disk material and the radiation are interacting
effectively, radiation pressure can act in that case
\citep{icke1989,fukue1996,chatto2004,vyas2015}.
On the other hand, if the cooling time scale is lower than
the infall time scale, matter can become
sufficiently hot and the required thermal pressure gradient force
can develop inside the accretion flow \citep[e.g., ][]{adios1999,soker2023}.
If an accretion disk,
having satisfied such conditions, additionally carries magnetic
field, all the effects may add up to make the outflow launching
favorable \citep[e.g., ][]{fukue2004}.

Outflows are not observed in isolation. Generally, spectral states
are correlated to the outflow activities for X-ray binaries as well
as active galactic nuclei (AGNs). Outflow is observed to be
enhanced during harder spectral states
\citep{fender2000,done2007,belloni2010,fender2012,ziarski2014}.
Thus, while focusing on
the theoretical understanding of the outflow formation, one should
pay attention to the fact that the same flow solution
has to produce a consistent spectral state.

There have been several analytical and simulation studies of
rotating, transonic accretion
flow, which is sub-Keplerian far away from the black hole
\citep{fukue1987,chakraba1989apj,cm1993,ryu1995apj,mrc1996,physrpt1996,
Das2014a,lee2016,kgbc2017,sukova2017,kgcb2019,gk2023,debnath2024}.
Being sub-Keplerian, the gravity is dominating far away. Hence,
the matter initially moves to the black
hole with nearly free-falling radial velocity and becomes
supersonic. However, for such a rotating accretion flow,
the outward centrifugal force can become comparable to the inward
gravitational pull closer to the black hole. As a result, the
free fall of matter is slowed down at a certain radius (few tens
to hundreds of Schwarzschild radii) and
the matter starts piling up. If the centrifugal force is
sufficiently strong, the flow can experience a shock. 
However, this mechanism differs from the boundary-layer shock model
\citep{Soker2003}, which applies to young stellar objects and involves 
shocks formed where the Keplerian disk interacts with the stellar surface. 
In contrast, our model concerns accretion onto black holes, where no 
physical surface exists, and the shocks arise dynamically from 
angular-momentum-induced centrifugal effects within the flow.
Such an accretion solution is shown to produce outflow and
outflow rate dependence on various inflow parameters are
studied analytically and numerically
\citep{mlc1994,rcm1997,chakraba1999,singh2011,ggc2012,aktar2015}.
The postshock region becomes hotter, denser and geometrically
thick. Bipolar outflow can originate from this region due to the combined
effects of centrifugal and pressure gradient forces. If the
accreting matter additionally brings in magnetic field, it
becomes stronger in the post-shock region due to compression and
contribute positively to the outflow launching
\citep{deb2017,bani2019,shende2019,mhd2020}.
The postshock flow becomes the base of the outflow.

This disk-jet solution simultaneously addresses the spectro-temporal
states. Increased density inside this postshock flow makes this region
optically thicker. As a result, it can intercept photons originating
within the accretion disk and Comptonize or inverse-Comptonize them.
In the popularly known two-component advective flow
\citep[TCAF,][]{ct1995,mondal2021} model, such
sub-Keplerian flow forms one component, and a Keplerian disk
component is sandwiched inside this. Lower energy seed photons
from the Keplerian disk is inverse-Comptonized to higher energy
by this post-shock region and produce harder spectrum. During
this time, outflow is also launched from the post-shock region
\citep{ggc2012, ggc2014}.
However, when the sub-Keplerian component is nearly absent, the
seed photons leave unaffected, producing a softer state and
outflow is also absent. In the case when the flow has single
component (i.e., only sub-Keplerian matter), even then, the seed
photons generated inside this component through the bremsstrahlung
or synchrotron process can get self-Comptonized
and produce harder radiations \citep{cm2006,mc2008}.
In such situations, outflow may accompany the harder state.

In this paper, we aim to investigate the fate of the bipolar outflow
launched from the postshock region, at a very large distance from
the black hole. We do not consider any magnetic field just to
focus on the effects of centrifugal and thermal pressure gradient forces
on the outflow. We are specifically interested to see whether
such a shocked accretion flow can launch sustained collimated
outflow reaching out to thousands of gravitational radii
and if yes, what terminal velocity
can they achieve? We present the results of several simulations with
increasing the specific angular momentum to explore the effect of
increasing centrifugal force on the outflow base (i.e., postshock
region) as well as on the outflow rate at far distance.
We additionally generate seed bremsstrahlung photons inside the disk-jet
system and compute the self-Comptonized spectra of these
seed photons using a monte-carlo based radiative transfer simulation
\citep{pss1983,gcl2009,ggcl2010,ggc2014}.

In this paper, we choose $r_g=2GM_{\rm{bh}}/c^2$ as the unit of distance,
$c$ as unit of velocity,
$r_g c$ as unit of angular momentum, and $r_g/c$ as unit of time,
unless specified otherwise.
Here, $G$ is the gravitational constant, $M_{\rm{bh}}$ is the mass of
the black hole, and $c$ is the speed of light in free space. 

\section{Simulation procedure} \label{sec:simul}

For our present study, we solve the inviscid hydrodynamics
equations and include thermal bremsstrahlung cooling in the
energy conservation equation as a source term. We assume
axisymmetry of the accretion disk. Thus, the system
can be described in two-dimensional cylindrical coordinates $R,Z$ as follows:

\begin{equation}
\frac{\partial \mathbf{U}}{\partial t} + \frac{1}{R}\frac{\partial \left(R\mathbf{F_R}\right)}{\partial R} + \frac{\partial \mathbf{F_Z}}{\partial Z} = \mathbf{S},
\label{eq1}
\end{equation}
where,
the vector of conserved variables $\mathbf{U}$, R-flux $\mathbf{F_R}$,
and Z-flux $\mathbf{F_Z}$ can be written as:
\begin{eqnarray}
{\bf U }= \left(
\begin{array}{ccccc}
\rho \\ \rho v_R \\ \rho l \\ \rho v_Z \\ E 
\end{array}
\right);
\quad
{\bf F_R}=\left(
\begin{array}{ccccc}
\rho v_R \\ \rho v_R^2 + P \\ \rho l v_R \\ \rho v_R v_Z\\ \left( E + P\right)v_R
\end{array}
\right);
%\end{eqnarray*}
\quad
{\bf F_Z}=\left(
\begin{array}{ccccc}
\rho v_Z \\ \rho v_R v_Z \\ \rho l v_Z \\ \rho v_Z^2 + P \\ \left(E + P\right)v_Z
\end{array}
\right)
\label{eq:flux}
\end{eqnarray}
And the source term is:
\begin{eqnarray}
{\bf S }&=& \left(
\begin{array}{ccccc}
0 \\ \frac{\rho v_\phi^2}{R} + \frac{P}{R}- \rho \frac{\partial \Phi}{\partial r}\frac{R}{r} \\ 0 \\
- \rho \frac{\partial \Phi}{\partial r}\frac{Z}{r} \\
- \rho \frac{\partial \Phi}{\partial r}\frac{\left(R v_R + Z v_Z\right)}{r} - Q_{br}
\end{array}
\right).
\label{eq:source}
\end{eqnarray}

Here, $\rho$ is mass density, $v_R, v_\phi, v_Z$ are the $R$, $\phi$ and $Z$ 
components of velocity, $P$ is pressure, 
$E=\frac{1}{2}\rho (v_R^2 + v_\phi^2 + v_Z^2) + \frac{P}{\gamma - 1}$ 
and $l=Rv_\phi$. $r$ represents the spherical
radius and is given by $r=\sqrt{R^2 + Z^2}$. $\Phi$ represents the
gravitational potential (\cite{pw1980}): $\Phi = - GM_{\rm{bh}}/(r - r_g)$.
We use a polytropic equation of state $P=K \rho^\gamma$ with $\gamma=4/3$.

Inside the simulation domain, $\rho$ is measured with respect to 
the reference density $\rho_0$ at the outer radial boundary:
\begin{equation}
\rho_0 = \frac{\dot{m}}{2\pi R_{\text{out}}\, h\, v_{R0}}
\label{eq:eq4}
\end{equation}
Here, $\dot{m}$, $h$ and $v_{R0}$ are the mass injection rate, thickness 
of the disk, and injection radial velocity, respectively, at $R=R_{\mathrm out}$.
In the chosen unit system, bremsstrahlung loss $Q_{br}$ is given as (\cite{msc1996})
\begin{equation}
Q_{\mathrm{br}} = 1.43\times 10^{-27}\,
	\rho^{2} \left(\frac{P}{\rho}\right)^{1/2}\,
\frac{\rho_{0} r_{g} T_{\mathrm{ref}}^{1/2}}{m_{p}^{2} c^{3}}
\label{eq:qbr}
\end{equation}
Here, $m_{p}$ is mass of a proton and $T_{\mathrm{ref}}$ is given by
\begin{equation}
T_{\mathrm{ref}} = \frac{\mu m_p c^2}{k_B}
\label{eq:eq6}
\end{equation}
$\mu$ = 0.5 and $k_B$ being the Boltzmann constant.

We solve the above equations using the code described in \cite{gk2023}.
We use the two-dimensional version of that code. It is a finite-volume
method-based globally second-order accurate code. van-Leer slope limiter 
\citep{mignone2014}
has been used for spatial reconstruction, HLL Riemann solver has been used
for interfacial flux calculation, and two-stage strong-stability preserving 
Runge-Kutta \citep{shu1988} method is used for time integration. For details, please
see the above references. The source terms are treated using an explicit method.

\subsection{Simulation set-up}

Our simulation set-up is very much similar to those which have been
used over decades to study such shocked accretion
flow \citep{cm1993,ryu1995apj,mrc1996,cam2004,otm2007,giri2010,om2012,
Das2014a,lee2016,kgbc2017,kgcb2019,gk2023,debnath2024}.
We perform our simulations on a vertically elongated $R-Z$ domain
$[0:250]\times[-2651:2651]$. All the simulations are run using
$480\times2000$ ratioed grid points having a common ratio $1.0043$.
Grids are constructed such that
we resolve the region close to the origin with the highest resolution
and grids become coarser as we move away. Thus, along the $R$ direction,
finer grids are located close to $R=0$ and along
the $ Z$ direction, ratioing is done symmetrically about the equatorial plane.
The black hole is placed at the origin, and we apply an absorbing 
boundary condition within a radius $r=1.8$.
Mass of the black hole is chosen to be
$M_{\rm{bh}}=6.5\times10^{9}$ M$_\odot$.
Mass injection rate $\dot{m}=0.01$ mass Eddington rate is assumed.
We present the results of a total of five simulations here.

\begin{table}[h]
\centering
    \begin{tabular}{c c c c c c c c p{5cm}}
     \hline
     
        ID & $l$  & $v_{R0}$ & $a_{R0}$ & Terminal velocity\\
        \hline
	    A1 & 1.50 & 0.03015 & 0.02907 & 0.05\\
	    A2 & 1.60 & 0.03006 & 0.02907 & 0.09\\
	    A3 & 1.70 & 0.02996 & 0.02907 & 0.12\\
	    A4 & 1.75 & 0.02991 & 0.02907 & 0.14\\
	    A5 & 1.80 & 0.02986 & 0.02908 & 0.1\\
        \hline
    \end{tabular}
    \caption{Simulation parameters where $l$, $v_{R0}$ and $a_{R0}$ 
	represent the specific angular momentum of the accretion flow, 
	radial velocity and sound speed of the injected matter respectively.
	Last column represents the terminal velocity of the outflow.
	See text for details.}
    \label{tab1}
\end{table}

Table \ref{tab1} shows the simulation parameters presented in 
this paper. Column 1 shows the case IDs. Column 2 shows the
specific angular momentum of the accretion flow $l$ at
the outer boundary. Columns 3 and 4 show the 
radial velocity $v_{R0}$ and the sound speed $a_{R0}$ of the injected
matter at the outer boundary. These are computed following the
theoretical calculation provided in \citet{chakraba1989apj}.
The calculation requires two conserved parameters, specific
energy 
$\epsilon=\frac{1}{2}(v_R^2 + v_\phi^2 + v_Z^2) + \frac{a^2}{\gamma -1} + \Phi$
($a=\frac{\gamma P}{\rho}$ being
the local sound speed) and specific angular momentum $l$ of the flow,
to figure out the radial variation of the flow profile.
For all our runs, we choose $\epsilon=0.001$. This $\epsilon$,
along with $l$ (Column 2), determines $v_{R0}$ and $a_{R0}$
at $R=R_{out}$.
Our choice of ($l, \epsilon$) is guided by the parameter
space classification of this flow, \citet{cd2001} which
shows that for these sets, shocks can
develop in the accretion solution. Additionally, we wish to investigate
the effect of increased centrifugal barrier
on the outflow and hence increase $l$ keeping $\epsilon$ constant.
The last column represents the terminal velocities attained by the 
outflowing matter in the respective simulation.
All simulations are run until a time stop of 160000. 
The solutions have achieved steady states by this time.

\subsection{Initial and boundary conditions}

We start our simulation with the entire domain initially filled with
static matter having constant floor density $\rho_{\mathrm {floor}}=10^{-6}$
and floor pressure
$P_{\mathrm {floor}}=\frac{a_{R0}^2}{\gamma}\rho_{\mathrm {floor}}$.
Accreting matter enters the simulation domain axisymmetrically
through the outer radial boundary $R=R_{\mathrm {out}}=250$. 
At this inflow boundary,
the disk is assumed to be thick, and the disk half-height is
$h/2 = 280$. Therefore, the inflow boundary condition is applied 
for $-h/2 \leq Z \leq h/2$ at $R=R_{\mathrm out}$. At the ghost 
grids adjacent to all the active grids within this $Z$ range,
we maintain time-independent matter density $\rho_0=1$, velocity
$v_{R0}$ (see Table \ref{tab1}) and pressure 
$P_0=a_{R0}^2 \rho_0/\gamma$ throughout the simulation
duration. The idea behind this is that the mass inflow rate towards
the black hole remains unchanged throughout the simulation duration.
At other $Z$ at $R=R_{\mathrm out}$, we apply outflow boundary
condition. Reflection boundary condition has been imposed on the
axis $R=0$. At the upper and lower $Z$ boundaries, outflow boundary
conditions are enforced.

\subsection{Bremsstrahlung emission modeling and self-Comptonization}
\label{brems}
Inside an accretion disk around a super-massive black hole, thermal
bremsstrahlung (free--free emission) constitutes an important 
radiative mechanism contributing to X-ray and $\gamma$-ray emissions. 
Since we perform simulations of geometrically thick accretion 
disks ($h(R)\sim R$, where $h$ is the accretion disk half-height at radius $R$) 
with small $\dot{m}$, we have low optical depth values 
(optical depth of 0.15 cm$^{-1}$ as shown in Section \ref{radiation})
Additionally, majority($\approx 99\%$) of the grid points in our simulation domain 
have temperature $> 2\times10^7 K $, hence we consider bremsstrahlung 
to be the dominant radiative mechanism and ignore line emissions.
The spectral emissivity (erg/cc/sec/Hz) due to bremsstrahlung emission
is given by \citep{rad_b}:
\begin{equation}
\frac{dW}{dV\,dt\,d\nu} = 
6.8 \times 10^{-38} \rho^2 T^{-1/2} \exp\left( -\frac{h\nu}{k_B T} \right)
\label{eq:emission}
\end{equation}
By integrating this expression over frequency, the total thermal
bremsstrahlung emissivity, as given in Equation~(\ref{eq:qbr}),
can be obtained.
These expressions form the foundation for modeling thermal 
bremsstrahlung emission in high-temperature plasma in the vicinity 
of black holes. 

To obtain the spatial distribution of photons, we use a Monte Carlo sampling
technique based on the cumulative distribution function (CDF)\citep{Sobol1994MC, ross2010}.
The total emission power (erg/sec) from each grid cell is first normalized to
construct a discrete probability distribution, where each element represents
the probability of a photon being emitted from the corresponding cell. 
The cumulative sum of this probability distribution refers to the CDF, 
which partitions the unit interval [0, 1] into segments proportional to 
the respective emission probabilities. A fixed number of random numbers are then 
drawn between [0, 1]. Each number is then mapped to the grid point corresponding 
to the CDF interval in which it falls, i.e., a random number smaller than the 
CDF value of the first cell is assigned to the first grid cell, while values 
ranging between two successive CDF values are assigned to the corresponding 
grid cell. In this way, the fraction of photons associated with each grid cell 
statistically reproduces the underlying emission probability distribution, 
with the accuracy improving as the number of sampled photons increases. 
The weight factor for modeling is calculated by dividing the integrated 
emission from each cell by the number of Monte Carlo photons distributed in that cell.

The seed photons for Comptonization are generated using the 
weighted photon distribution described above, followed by modeling 
the energy distribution through Gamma function modeling
\citep{sobol, gamma_model} using a single random number.

Once a photon is generated within the disk-outflow configuration, following
the above algorithm, we track the photon until it leaves the entire 
computational domain, or it is absorbed by the black hole. 
During its tracking, if it satisfies a scattering 
condition (i.e., its optical depth becomes larger than a threshold
value), a (inverse-)Comptonization is modeled. For modeling
these tracking and Comptonization processes, we follow the same methods
used in \citet{1983ASPRv...2..189P,gcl2009, ggcl2010,ggc2012,ggc2014}.
We also consider gravitational red-shift during the tracking
of the photons.
The spectrum is calculated in a post-processing manner on the final 
disk-outflow configuration.
Thus, we do not include the heating/cooling effects due to the
Comptonization part in the time-dependent fluid dynamics.
For our current model parameters, we observe that a very small fraction
($< 1\%$) is intercepted by the disk-outflow
configuration. Thus, we believe this exclusion of heating/cooling effects
won't affect the fluid dynamics significantly.
We use 10$^8$ seed photons for the spectral modeling. This number has
been chosen based on a convergence test. We ensure that the final
spectrum is converged and Monte Carlo noise is upto the level
of tolerance.

\section{Results} \label{sec:results}

\subsection{Flow dynamics}

\begin{figure}[ht!]
    
    \includegraphics[width=0.8\textwidth]{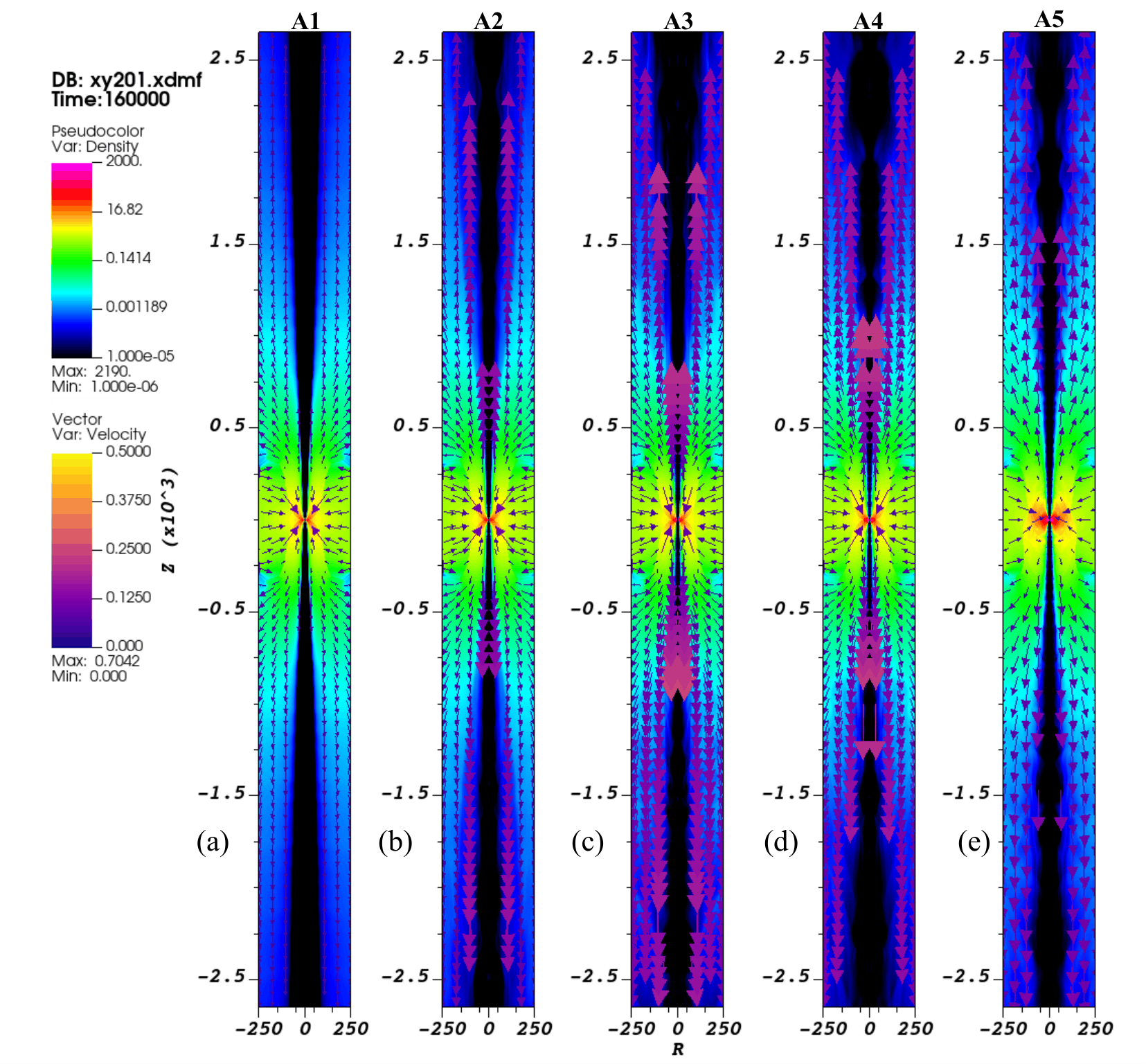} 
\caption{ shows the mass density ($\rho$) distribution (in color) on log scale 
overlaid with velocity
    arrows at the final time for all cases 
    A1-A5 (a:A1, b:A2, c:A3, d:A4, e:A5). Black color shows lower 
    densities, and red color shows higher densities. Length of a velocity
	arrow is proportional to the magnitude $\sqrt{v_R^2+v_Z^2}$. Colors of arrows
	represent their magnitudes as shown in the legend for velocity vector.
	The maximum velocity is found extremely close to the black hole on the equatorial plane
    }
    \label{fig:fig1}
    
\end{figure}

Fig. ~\ref{fig:fig1} shows the density-velocity distribution at the
final time for all the runs A1-A5 (a:A1, b:A2, c:A3, d:A4, e:A5). 
Colors represent the mass density on a log scale and the velocity field
($v_R, v_Z$) are overplotted on this as arrows. Black color shows 
lower density ($\approx 10^{-5}$ w.r.t. reference density $\rho_0$)
and red color shows higher density. 
Length of an arrow is proportional to the velocity magnitude
$v_r=\sqrt{v_R^2+v_Z^2}$.
Although the simulations are conducted inside the radial domain
[$0:R_{\mathrm out}$], we have reflected results about the vertical axis while plotting.
Matter enters the simulation domain at $R_{\rm out}=250$ with the
velocity vectors pointed towards the black hole located at the center.
After the transient phase is over (by time $t \sim 20000$), the simulation
domain is mostly filled with disk-outflow matter. We observe that
the region near the equator is filled with a geometrically thick
accretion disk, while regions vertically up and down are filled 
with outflowing matter. The outflow is hollow in structure, meaning
the region along the axis is void of matter.
\begin{figure}[ht!]
    
    \includegraphics[width=1\textwidth]{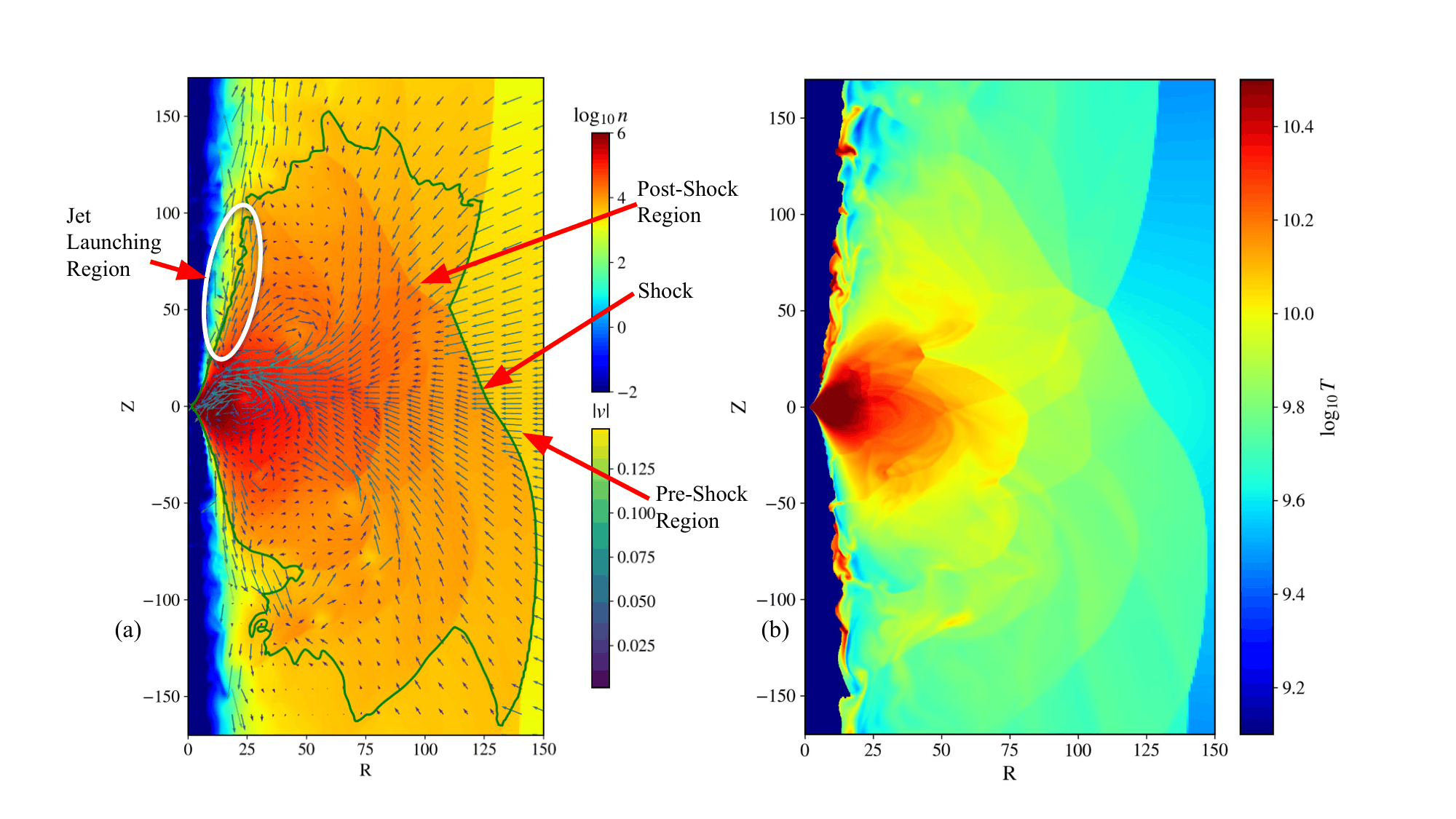} 
    
 	\caption{(a) shows a zoomed-in view of case A5 
	in the $R-Z$ domain $[0:150]\times[-160:160]$. 
	We plot the number density (measured in particle number/cc) colormap overlaid with 
	velocity arrows. The iso-density contour (green line) corresponding to the
	value $8.85\times10^6$, allows to identify the shock surface at $R\sim 100$.
	The region inside this contour is the high-density post-shock region, 
	and the outer low-density region is the pre-shock region, marked accordingly. 
	The post-shock region shows the presence of turbulent vortices.
	We also mark the jet launching region close to the black hole. 
	(b) This figure shows the temperature (in Kelvin) distribution in the same
	domain clearly separating the high-temperature post-shock region from 
	the low-temperature pre-shock region for case A5.
    }
    \label{fig:fig2}
\end{figure}

Zoomed-in observation of the equatorial region reveals the thick disk
structure as well as the outflow launching mechanism. Fig ~\ref{fig:fig2} shows
the central region $[0:150]\times[-160:160]$ for the case A5. 
Fig ~\ref{fig:fig2}(a) shows the number density distribution and (b) shows
the temperature distribution. We notice
that as the infalling matter moves towards the central region,
density (and temperature as well) increases due to geometric compression
initially and then due to a shock discontinuity. The discontinuity
structure bends outward away from the equatorial plane \citep{mlc1994,mrc1996}
giving rise to a centrifugal pressure-supported
boundary layer (CENBOL). Matter density increases past this
discontinuity surface again due to geometric compression. An iso-density
contour plot may allow one to see the geometrically thick, toroidal disk structure
in this region.
The green iso-density contour corresponding to number density value $8.85\times10^6$
in Fig ~\ref{fig:fig2}(a) shows such a toroidal structure.
Velocity vectors in the post-shock region show presence of turbulent vortices as well.
We also notice from the velocity vectors that outflow is primarily launched from
this post-shock region. This observation holds for all the cases.
However, the centrifugal force varies from one case to another
and that affects the shock strength and, hence, the compression.
This causes the results to differ. We provide further analysis below
to quantify these.

\begin{figure}[ht!]
    \centering
    \includegraphics[width=\textwidth]{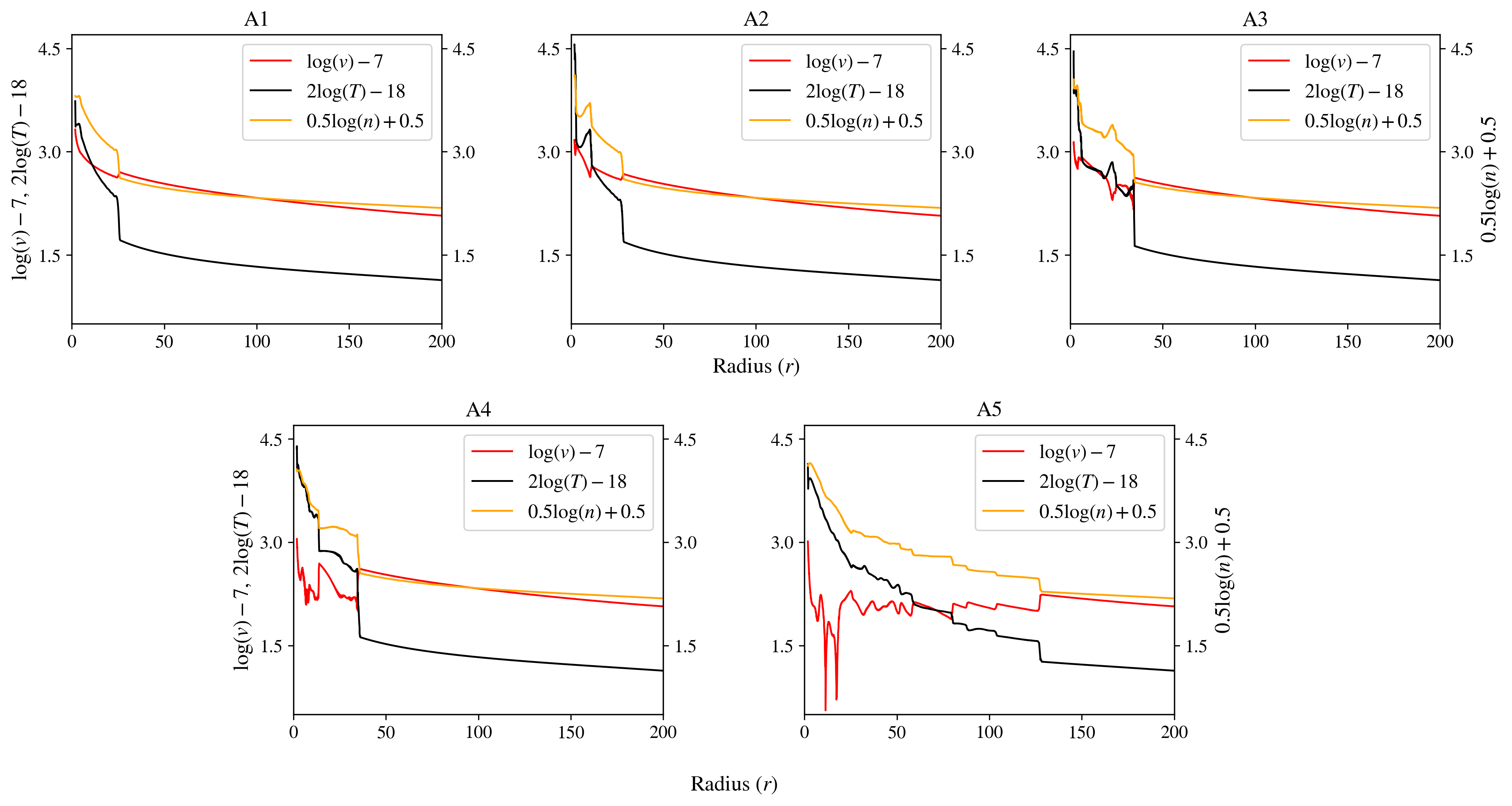}
    \caption{ represents the radial variations (along the equator) 
	of logarithm of temperature($T$), number density ($n$), and 
	velocity ($v_R$) at the final time. Different quantities
	are translated by a scalar along the vertical axis to bring them to the same
	scale. We show temperature in Kelvin, number density in particle number/cc and velocity in cm/s units.}
    \label{fig:fig3}
\end{figure}
The discontinuous jump and subsequent compression become prominent
when we plot the radial variation of different fluid variables.
Figure~\ref {fig:fig3} shows the radial variation of number density, velocity,
and temperature along the equatorial line for all the cases at the final
time. Density and temperature are measured in particle number/cc
and Kelvin units, respectively. 
For the low angular momentum cases ($\lambda = 1.50$ 
and $1.60$, corresponding to runs A1 and A2 respectively), the flow shows
a mild deceleration (velocity drop) at the centrifugal barrier
around $R=28 $ (A1) and $R=32 $ (A2). A density compression (compression ratio
$\rho_{\rm post-shock}/\rho_{\rm pre-shock}\sim 6.25$(A1) and $\sim 6.28$(A2))
and temperature enhancement is seen due to this. For 
slightly higher angular momentum $\lambda = 1.70$ (run A3), the 
centrifugal barrier becomes strong enough to produce a well-defined 
shock (supersonic to subsonic transition) at $R=34.5 $,
resulting in a sharp, larger drop in velocity. This causes larger density
compression and temperature rise. The density compression ratio is found to
be $7.4$ in this case. We also observe the presence of an
inner shock near $R=5$. For $\lambda = 1.75$ (run A4), the 
shock feature is more pronounced (shock at $R=34.8$), and 
the density compression ratio is found to be $8.7$. The flow also exhibits
evidence of a secondary, inner shock closer to the black hole
($R=13.5 $) before passing through the inner sonic point. For the 
highest angular momentum $\lambda = 1.80$ (run A5) case, the flow shows
unsteady behavior, both radial and vertical oscillations of the
post-shock matter are observed. Careful observation of successive snapshots show
the formation and disappearance of turbulent vortices of different
sizes in the post-shock matter. Since the centrifugal barrier is very
strong in this case, the bounced-back matter from this barrier interacts
with the incoming matter, resulting in the initial formation of large vortices,
above and below the equator. Subsequent velocity shear between the
the incoming flow and the matter in these vortices develop shear 
instability. This instability possibly grows in time, resulting in
the formation of turbulent flow. Thus, the post-shock region 
becomes highly turbulent with many intermediate discontinuities. 
The radial profiles of velocity, number density, and temperature at the final
time for this case show such discontinuities. Such unsteady shock 
dynamics for high angular momentum flow have been reported in earlier 
numerical simulations \citep{rcm1997,deb2016,gk2023}.

\begin{figure}[ht!]

 \centering
\includegraphics[width=0.8\textwidth]{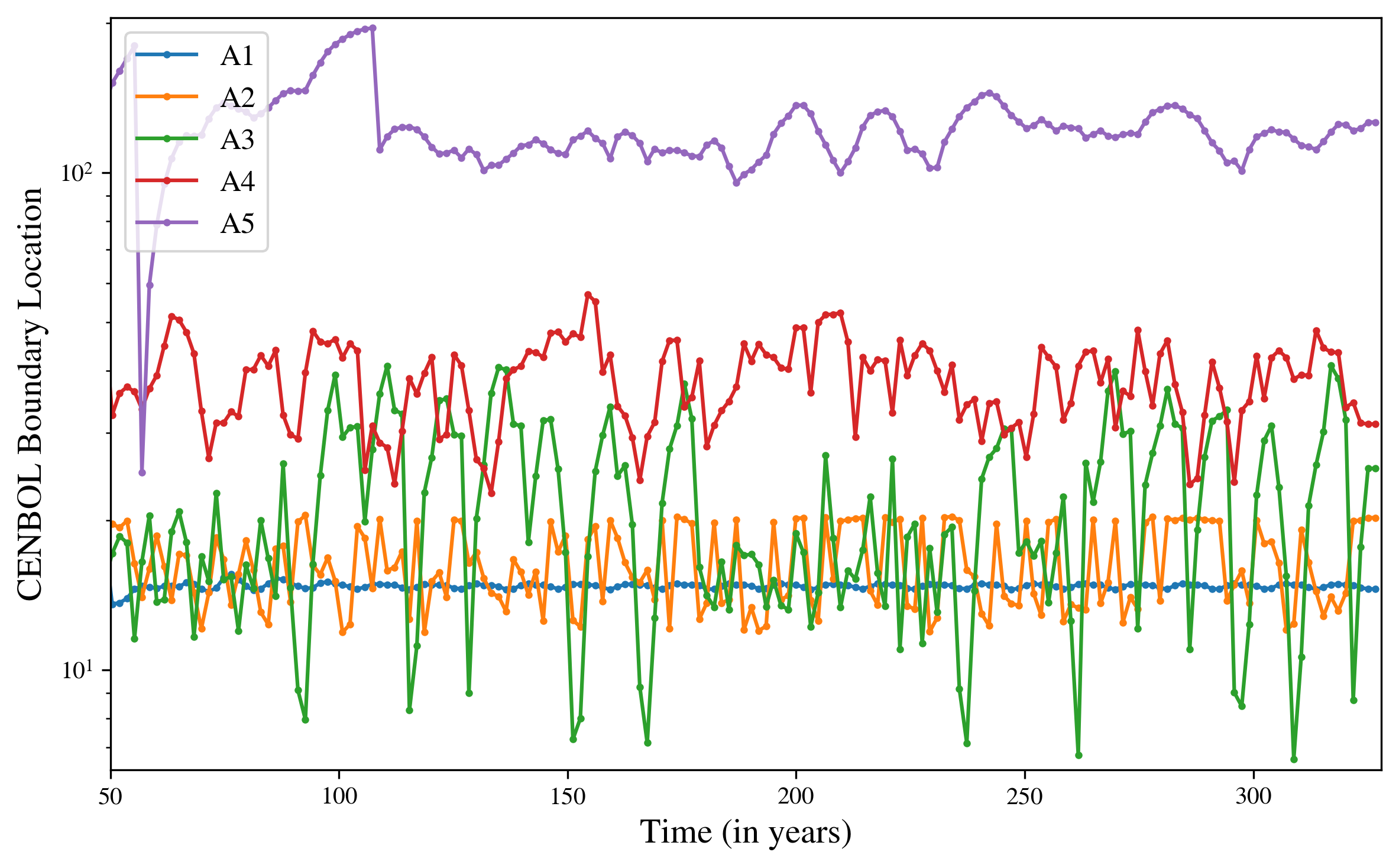}

\caption{shows temporal variation of the CENBOL boundary at the equator for 
all cases. Cases A1 and A2 do not show shock formation, though a discontinuity
due to centrifugal barrier is found. Cases A3, A4, and A5 show 
shock formation and oscillations of the CENBOL boundary over time. }
\label{fig:fig4}
\end{figure}

The CENBOL structure is nearly stationary for the low angular momentum
runs (A1 and A2), while cases A3 and A4 exhibit quasi-periodic
radial oscillations associated with shock transitions. 
For the turbulent, high angular momentum case (A5), the boundary layer
is highly dynamic, showing significant temporal variability. 
In Fig. \ref{fig:fig4}, we plot the time variations of the CENBOL 
at the equator for all five cases. Time is represented in years here. 
With increasing centrifugal force, the average boundary location moves
farther from the black hole. 

\subsection{Outflow}

The dynamical CENBOL is found to launch outflow from a region close
to the black hole. At the boundary layer, flow is slowed down, and as
a result, the kinetic energy of the flow is converted to thermal energy.
Thus, inside the CENBOL, the flow becomes hotter. Further geometric
compression makes the rotating flow even hotter. Thus, the flow 
develops a pressure gradient in both radial and vertical directions. 
This vertical pressure gradient force launches outflow of matter 
in the vertical direction. Since the ram pressure of the 
infalling matter prevents matter from expanding in the radial direction, 
there is no outflow in the radial direction. Thus, we see only bipolar
outflow being launched in all the cases.

\begin{figure}[ht!]
    \centering
    \begin{subfigure}[t]{0.48\textwidth}
        \centering
        \includegraphics[width=1.15\textwidth]{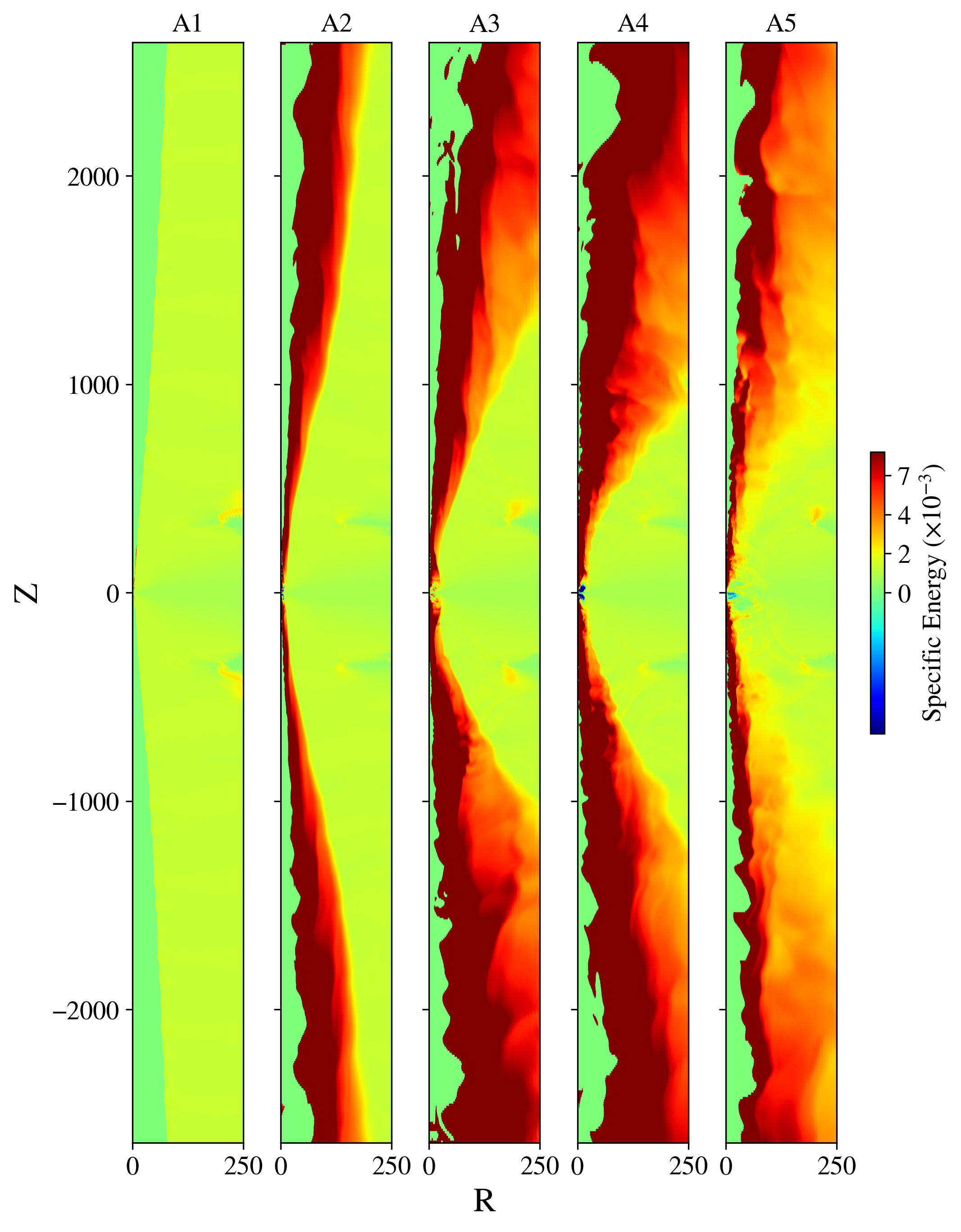}
        \caption{}
        \label{fig:fig5a}
    \end{subfigure}
    \hfill
    \begin{subfigure}[t]{0.48\textwidth}
        \centering
        \includegraphics[width=0.68\textwidth]{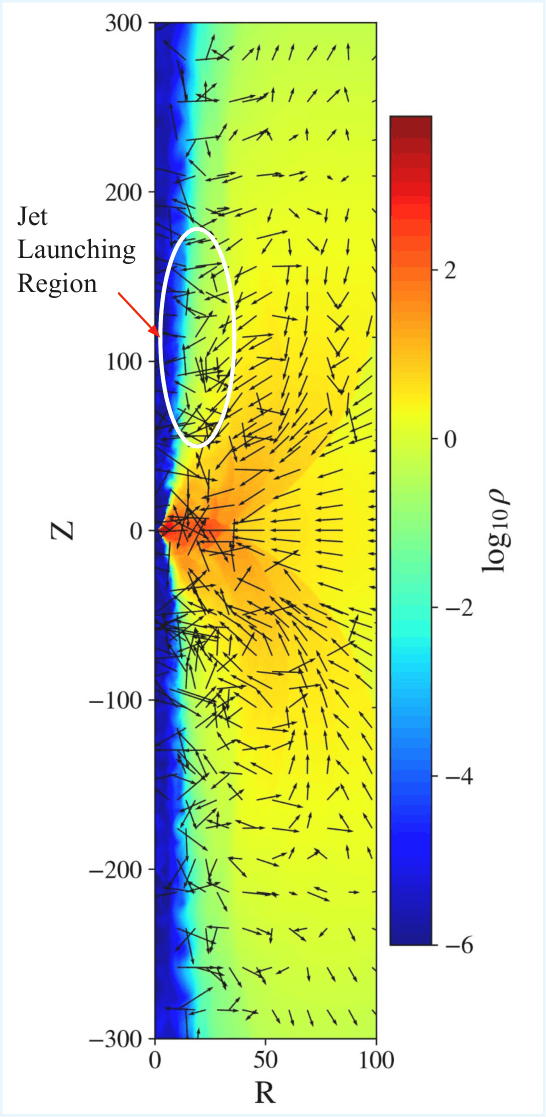}
        \caption{}
        \label{fig:fig5b}
    \end{subfigure}

\caption{(a) shows a colormap of specific energy ($\epsilon$) distribution 
    for all cases A1-A5 at the final time. The high-energy 
    matter is shown by red color, which forms the outflow.
    (b) shows the force vectors overlaid on the
    log$_{10}\rho$ colormap for case A3, clearly showing 
unidirectional force 
    vectors beyond $-200$ and $200$ in the Z-direction. Force is calculated in units of $\frac{c^2}{r_g}.$}
\end{figure}

To investigate whether the outflowing matter can actually become unbounded,
we plot the specific energy($\epsilon$) distribution
inside the disk-outflow region in Fig. \ref{fig:fig5a}.
These plots have been done using the final timestep data. For all the runs,
infalling matter has $\epsilon=10^{-3}$. This value is mostly 
conserved within the accretion disk of thickness $-280<Z<280$
at all radii except very close to the black hole. Inside the compressed region,
energy is redistributed because of the turbulent interaction between
the incoming and the centrifugal barrier reflected matter. Thus,
we observe some matter even with a negative $\epsilon$. Such matter
will be ultimately absorbed by the black hole. Importantly, we observe
that the outflowing matter generally has higher values of the $\epsilon$ and is always positive, making sure that
this matter will contribute to the unbound outflow.

\begin{figure}[ht!]
    \centering
    \includegraphics[width=0.8\textwidth]{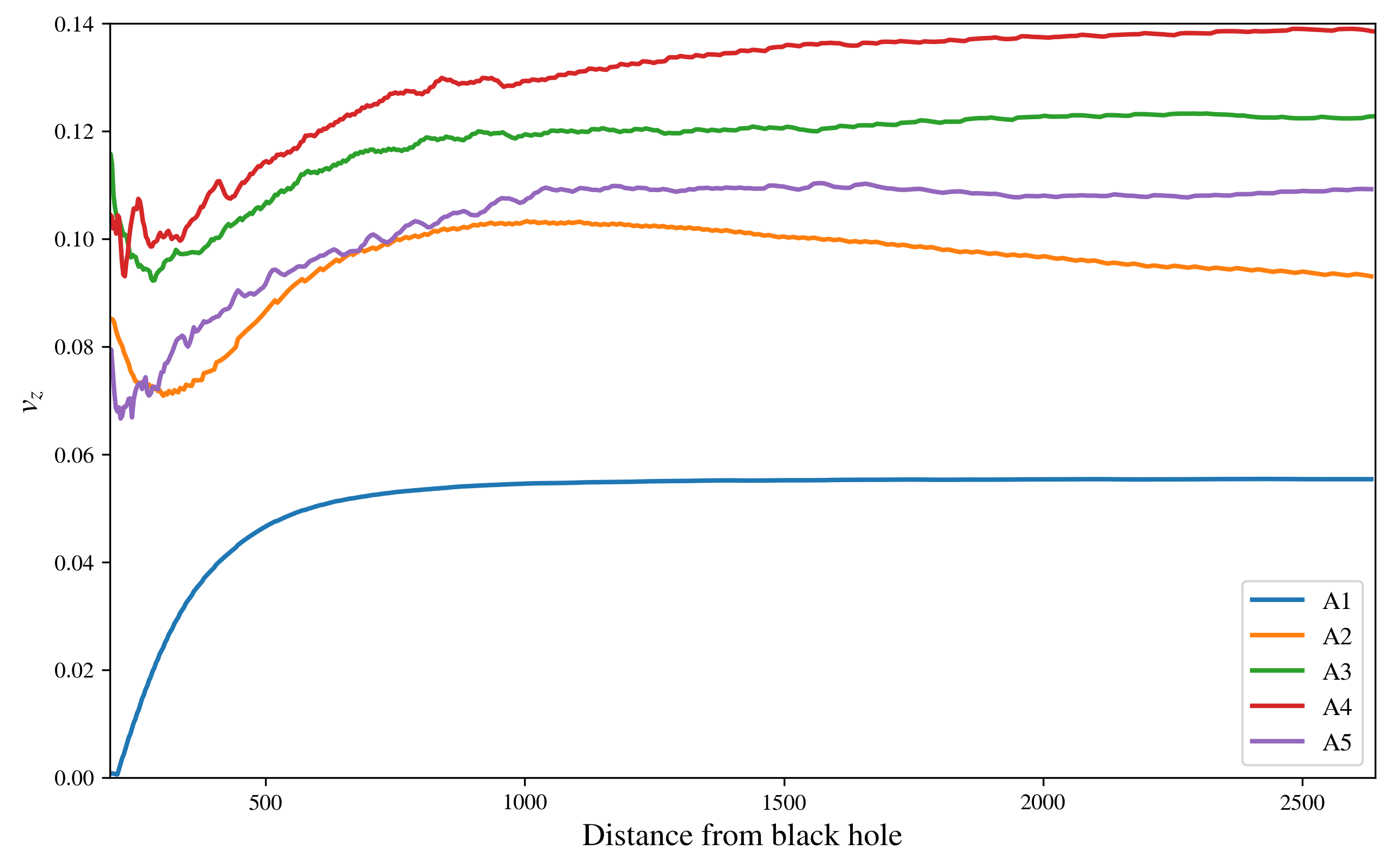}
    \caption{shows time-averaged outflow velocity as a function of 
radial distance from the origin for all the cases. Outflowing
matter is found to attain a terminal velocity after it moves a certain
distance from the black hole.}
    \label{fig:fig6}
\end{figure}

Fig. ~\ref{fig:fig6} shows the time-averaged vertical velocity
($v_z$) of the outflowing matter as a function of radial distance from 
the central black hole for different cases.
The time averaging has been performed over the entire steady-state duration. 
From velocity vectors in Fig. ~\ref{fig:fig1}, we note that within 
$|Z|\lesssim 300$, the outflowing region has a thin width. The velocity
shear between outflowing and inflowing matter in this region
develops Kelvin-Helmholtz instability. Thus, we do not get reliable values
of $v_z$ in this region. As matter moves away beyond $|Z| \sim 300$,
the outflow expands radially as the vertical disk thickness of the inflowing 
accreting matter is less than 300. The outflow is mostly hollow in structure, 
and the inner surface is highly dynamic. Thus, while calculating 
the $v_z$ at a given distance, we ensure that the distant point is well 
inside the outflow region.

From Fig. ~\ref{fig:fig6}, we observe that outflowing matter accelerates
initially and attains a terminal velocity as it moves away.
The initial acceleration to the outflowing matter is provided by
the pressure gradient and centrifugal forces.
We show the net force distribution (force vectors overlaid on log$_{10}\rho$
distribution) around the central region for case A3
in Fig. ~\ref{fig:fig5b}. This plot reveals that the net force 
below $|Z|\lesssim200$ is mostly inward. However, a very thin layer of
matter close to the Z-axis (marked by white oval) 
experiences outward force. Because of the K-H instability in the 
region, it is very difficult to clearly identify
unidirectional outward force. For $|Z| > 200$, we start observing
outward directed force vectors. This force imparts initial acceleration
to the outflowing matter.
However, further acceleration stops as matter moves away from the
denser accretion disk. In the absence of any force away from the
black hole, it attains a terminal velocity. The net force at the outflow
base accelerates the matter beyond the escape velocity so that
it can escape the gravitational pull of the central black hole.
For low angular momentum runs (A1 and A2), the terminal velocity
of the outflow remains small, saturating at $v_z \lesssim 0.09$.
For these cases, density and pressure compressions at the 
discontinuity are lower, as can be seen from Fig. ~\ref{fig:fig3}. 
Also, the centrifugal force is less due to lower values of $l$. 
For intermediate values ($\lambda = 1.70$ run A3), the outflow 
velocity increases more efficiently, reaching $v_z \approx 0.12$. 
Here, the compressions and centrifugal force are higher than in 
the previous two cases. The maximum terminal velocity is obtained 
for $\lambda = 1.75$ (run A4), where $v_z$ saturates near $0.14$. 
We observe the highest compressions in this case. For the highest 
angular momentum run ($\lambda = 1.80$ run A5), the terminal 
velocity is significantly reduced ($\sim 0.1$). In this case, 
although the centrifugal force is highest, the compression is diluted 
due to several shocks and turbulence formations.

\begin{figure}[ht!]
    \centering
    \includegraphics[width=0.8\textwidth]{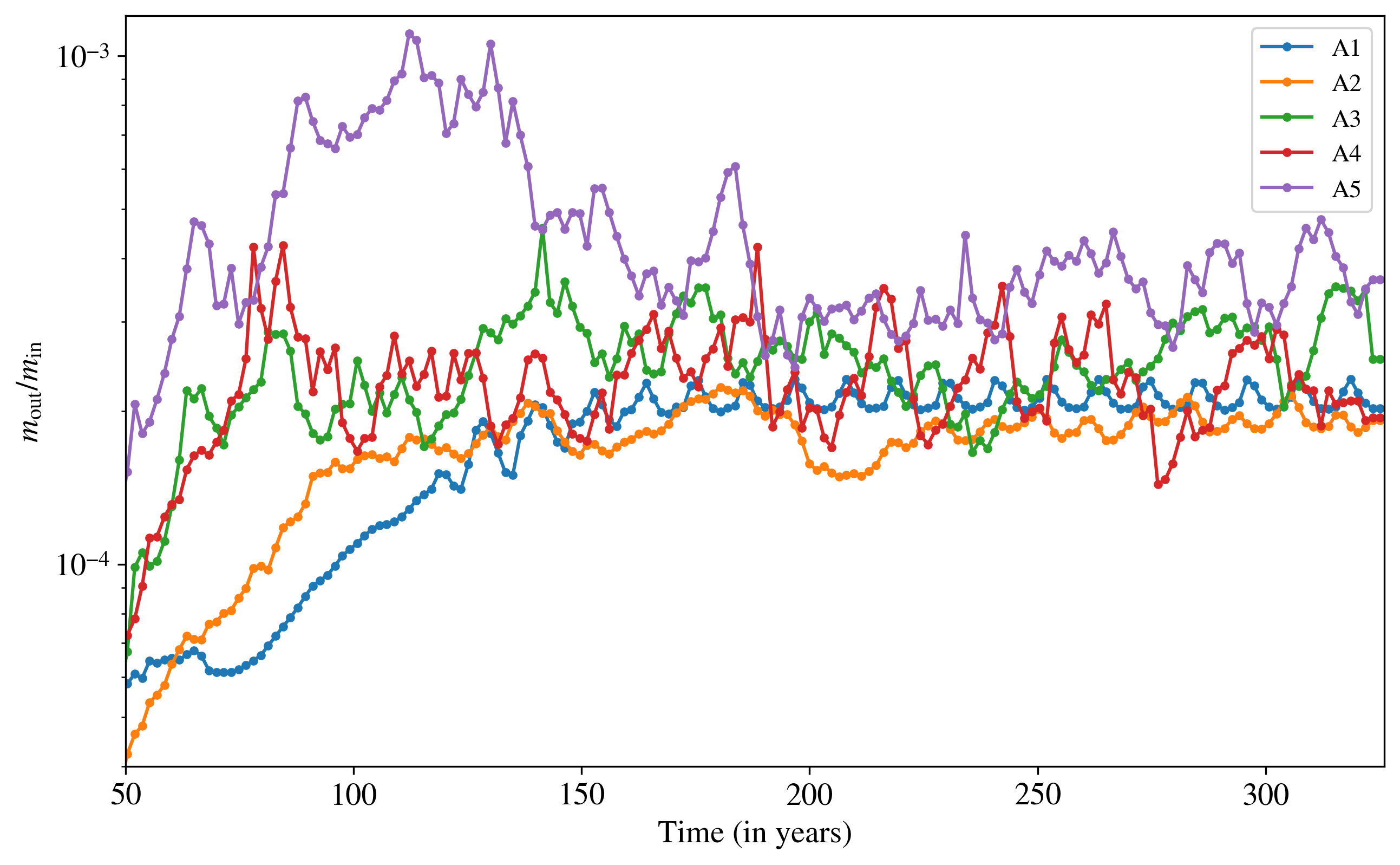}
    \caption{shows the outflow rate, normalized by the respective injection
rate, over time.}
    \label{fig:fig7}
    
\end{figure}

We also quantify the outward mass flux through the top and the bottom
surfaces ($|Z|=2651$) of the computational domain. Fig. ~\ref{fig:fig7}
shows the total mass flux through these two surfaces, normalized
by the mass injection through the outer radial boundary. On average,
we find the highest mass outflow rate ($\sim$ 0.04 \% of the injection
rate) for the highest $l$ (run A5) case. We emphasize that the
measured outflow rate is within $\sim$ 5$^\circ$ opening angle of the
jet.

\subsection{Radiative properties}
\label{radiation}

\vspace{0.2cm}
\begin{figure}[ht!] 
\centering
\includegraphics[width=0.8\textwidth]{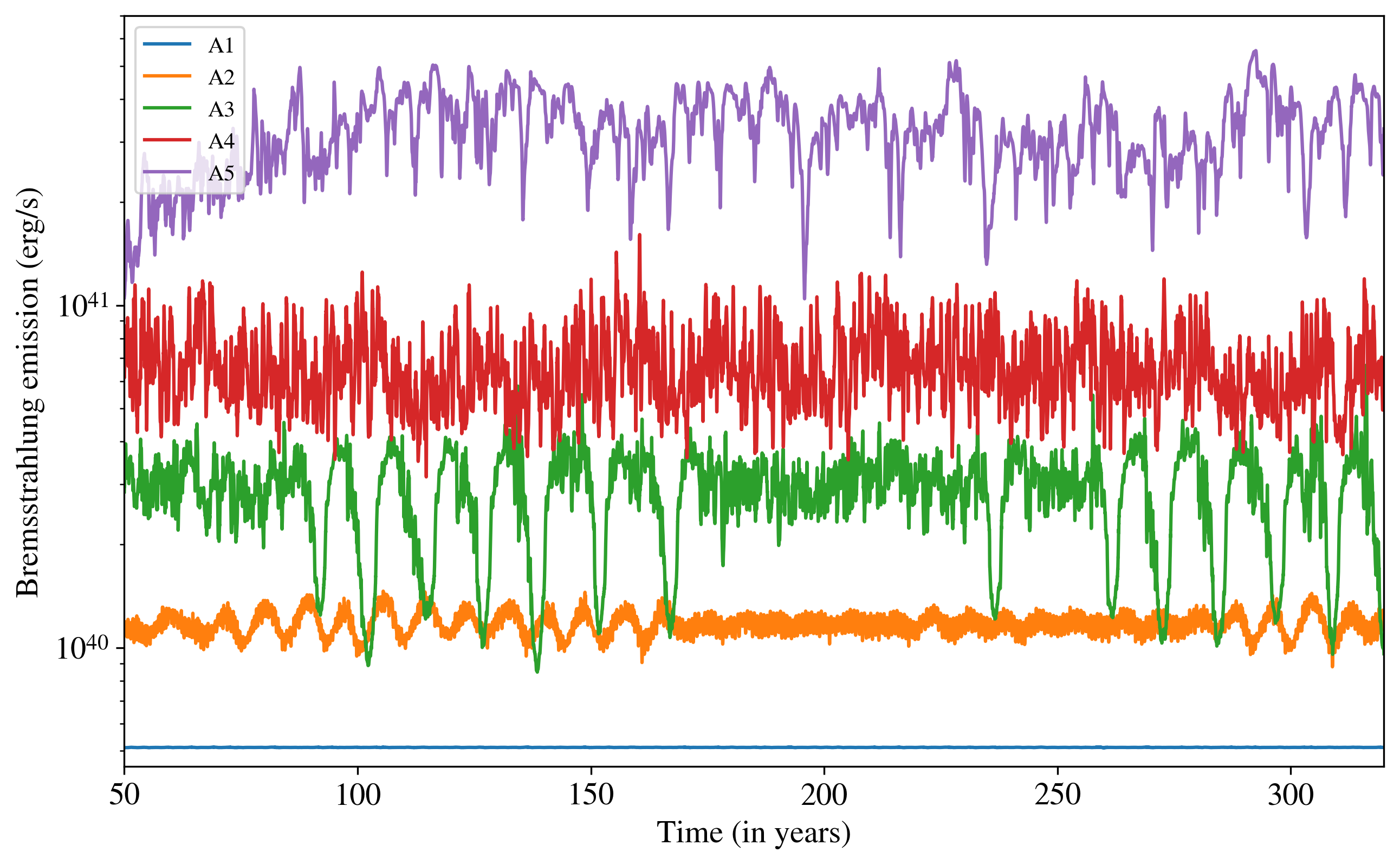}
\caption{shows the bolometric bremsstrahlung luminosity as a function 
of time for all accretion cases A1–A5. All our simulations are done
for a black hole of mass of $M = 6.5 \times 10^9 \, M_\odot$ with 
a mass accretion rate of $\dot{m} = 0.01$ (in units of the mass 
Eddington rate), corresponding to a physical mass accretion rate 
of $\dot{M} \approx 9.36 \times 10^{24} \, \mathrm{g\,s^{-1}}$ 
($\approx 0.15 \, M_\odot \, \mathrm{yr^{-1}}$). 
The resulting bolometric luminosity is of the order of $10^{42}$ 
for the case A5 \, $\mathrm{erg\,s^{-1}}$, yielding a radiative 
efficiency of $L_{\mathrm{bol}}/(\dot{M}c^2) \approx 4.29 \times 10^{-4}$.}
\label{fig:fig8}
\end{figure}
\begin{figure}[ht!]
\centering
\includegraphics[width=0.9\textwidth]{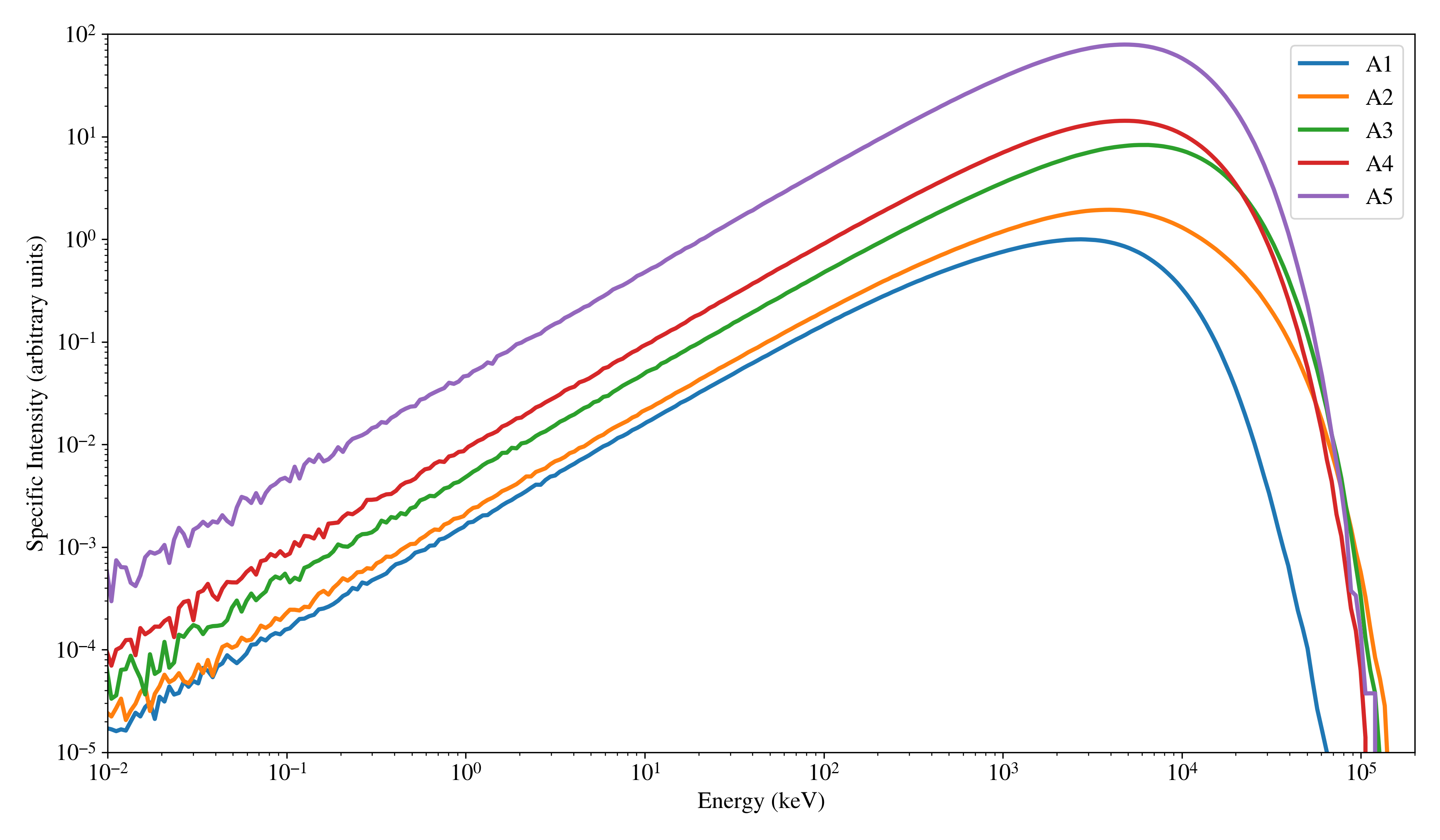}
\caption{shows the net (seed plus the Comptonized) bremsstrahlung spectra for
all the cases. See text for more details.}
\label{fig:fig9}
\end{figure}

Fig. ~\ref{fig:fig8} shows the temporal evolution of the
bolometric bremsstrahlung luminosity (erg/sec)for the five cases marked in the
legend. Time is measured in years. The bremsstrahlung 
emission is computed as the volume-integrated
emissivity of the plasma using Eq. \ref{eq:qbr} and
primarily contributed by the density
and temperature distribution inside the CENBOL region.
For the lowest angular momentum case A1 ($l = 1.50$), the emission
remains very low and nearly constant throughout the simulation.
This is consistent with the absence of strong compression and
variability inside the CENBOL region as found in Fig. ~\ref{fig:fig3}
and Fig. ~\ref{fig:fig4}. 
For case A2 ($l = 1.60$), the luminosity rises slightly and shows small
amplitude variability compared to A1.
For higher $l$ cases (runs A3, A4, and A5), we observe the bolometric
luminosity to increase with $l$. At the same time,
all these cases show significant variability; the amplitudes
vary by a few factors.
Average size of the CENBOL increases with increasing $l$. Thus,
increasingly larger volumes with denser matter enhance the
bremsstrahlung emission. The variability is caused by the 
dynamical CENBOL. Fig. \ref{fig:fig4} already
indicates the variability of CENBOL, and observed variability in
luminosity is consistent with this. 

We also investigate the spectral properties of the disk
due to this bremsstrahlung emission and its self-Comptonization effect.
We model the bremsstrahlung emission and (inverse-)Comptonization of the
emitted photons using Monte-Carlo techniques as described above in 
Section \ref{brems}. Fig. \ref{fig:fig9} shows the final spectra
for all the cases. The plots show the specific intensity (in arbitrary units)
as a function of photon energy (measured in keV). Since we are
performing single-component fluid dynamics (i.e., electrons or
protons/ions are not separated), we have a single-component temperature.
Fig. \ref{fig:fig3} shows the radial profile of this single-component 
temperature along the equator. The temperature
profiles of the fluid result in the bremsstrahlung spectra to
have an exponential cut-off at very high energy, and hence, the
specific intensities peak close to 5000 keV. We have chosen a low energy
cut off at 0.01 keV in the spectra because the 
specific intensity falls off by
an order of 5 below $0.01$ keV. Because of the lower density, 
most of the photons escape without being scattered.
Among all the cases, we find the highest scattering events for case $A5$
around $0.25$\% of the seed photons got scattered. 
Highest optical depth is found to be 0.15 cm$^{-1}$ for this case 
integrated along the equatorial plane(edge-on). For all the cases,
the highest energy photons originate from the innermost high-temperature 
region. As these photons travel away from the black hole,
they mostly get redshifted. Also, we find that if such high-energy 
photons get
scattered at the outer, cooler region, they mostly deposit their
energy to the electrons via Comptonization.

In passing, we would like to mention that even though our disk-jet
systems produce photons of energy $\sim$ MeV, we do not consider the 
electron-positron pair production. The possibility of pair production in 
high-energy astrophysical sources is not determined solely by photon energy. 
Instead, it depends critically on the radiation density, which is quantified 
by a dimensionless parameter \citep{gfr1983}:
\begin{equation}
\ell = \frac{L \sigma_T}{D m_e c^3},
\end{equation}
where $L$ is the luminosity measured in units of Eddington luminosity,
$D$ is the characteristic size (measured in $r_g$ unit) of the 
emitting region, $\sigma_T$ is the Thomson cross-section, and $m_e$ is the electron mass.
For our case, the Eddington luminosity is
\begin{equation}
L_{\rm Edd} = 1.3 \times 10^{38} \left(\frac{M_{bh}}{M_\odot}\right)
\simeq 8.45 \times 10^{47} \ \mathrm{erg \ s^{-1}}.
\end{equation}
The maximum bolometric luminosity we observe from the result 
$L_{\rm bol} \sim 10^{42} \ \mathrm{erg \ s^{-1}}$.
Hence, $\frac{L_{\rm bol}}{L_{\rm Edd}} \sim 10^{-6}
$, if we consider 
$D \sim 2651 r_g $,
we find $
\ell \sim 10^{-9}.$
Since $\ell \ll 1$ \citep{1984MNRAS.209..175S}, the photon--photon optical 
depth is extremely small. Although the bremsstrahlung spectrum extends up to 
$\sim 100$ MeV, the radiation field is too dilute for significant 
pair production. High-energy photons escape without generating a pair cascade.

\section{Summary and Conclusions}
\label{discuss}

In this paper, we study unmagnetized accretion disk-outflow system around
a super-massive black hole. These simulations results are equally 
applicable for stellar-mass black holes, also provided the radiative 
efficiency is low. We simulate the accretion of nearly free-falling
sub-Keplerian matter which develops a centrifugal barrier due to
non-zero angular momentum. A part of the infalling matter is
compressed due to this barrier and leaves the system as outflow. 
In this work, we investigate this mechanism and quantify the outflow
by varying the angular momentum of the infalling matter. We
use radiative-hydrodynamics simulations for our study, where
radiative cooling due to bremsstrahlung emission is included.
Our main findings and conclusions are as follows:
\begin{itemize}
\item Nearly free-falling sub-Keplerian matter slows down close
to a black hole due to a strong centrifugal barrier. The strength
of the barrier depends upon the specific angular momentum of the flow.
Because of the barrier, a shock develops in the flow and shock
compressed flow forms a hotter, denser corona surrounding the
black hole.
\item The shock-compressed flow subsequently accelerates towards the
black hole and further gets geometrically compressed. Part of this 
compressed, rotating flow experiences a net outward force due
to the combined effects of pressure gradient and centrifugal forces,
which accelerates flow away from the black hole.
\item This accelerated matter becomes unbounded and escapes
the accretion disk as a collimated, bipolar outflow out to thousands
of radii. In the absence of further
acceleration, this outflowing matter achieves a terminal speed 
that depends on the strength of the centrifugal barrier and the resulting
compression.
For our simulated cases, we find a maximum terminal speed of $0.14c$. 
Net mass flux of outflow also depends on the specific angular momentum.
\item The post-shock flow shows dynamical behavior that reflects
upon the outflow and outgoing radiative properties from the accretion
disk. The mass flux in the bipolar outflow and the total luminosity show
significant temporal variabilities consistent with the variability
seen in the CENBOL boundary at the equator.
\item The bremsstrahlung spectra are primarily determined
by the innermost, dense region of the accretion flow. This makes
the resulting spectra peak at high energies and produce hard
spectral states. Low optical
depth of the accretion disk, resulting due to low accretion rate,
doesn't show a significant Comptonization effect.
  
\end{itemize}

\normalem
\begin{acknowledgements}
We acknowledge the usage of the Kepler Computing facility,
maintained by the Department of Physical Sciences, IISER Kolkata.
We also acknowledge the usage of IUCAA's Pegasus Computing facility
for conducting a few simulations.
Authors also acknowledge the help of Mr. Bungkiu Kissinquinker
during the initial phase of this work.
\end{acknowledgements}

\bibliographystyle{raa}
\bibliography{ref}

\end{document}